\documentclass[12pt]{sn-jnl}
\usepackage{amssymb, amsthm}
\usepackage{amsmath}
\usepackage{mathtools}
\usepackage{algorithm}
\usepackage{algpseudocode}
\usepackage{geometry}
\usepackage{fullpage}
\usepackage{graphicx}
\usepackage{booktabs}
\usepackage{array}
\usepackage{cite}
\usepackage{hyperref}   
\usepackage{enumitem}
\usepackage{fullpage}
\title{HexaMorphHash \(H\textsubscript{M}H\) - Homomorphic Hashing for Secure and Efficient Cryptographic Operations in Data Integrity Verification}

\author[1]{\fnm{Mr.} \fnm{Krishnendu} \sur{Das}\thanks{Currently leading research in homomorphic encryption and secure multi-party computation.}}

\affil[1]{\orgdiv{DST-CIMS}, \orgname{Banaras Hindu University}, \orgaddress{\city{Varanasi}, \country{India}}}

\begin{document}

\abstract{
In the realm of big data and cloud computing, distributed systems are tasked with proficiently managing, storing, and validating extensive datasets across numerous nodes, all while maintaining robust data integrity. Conventional hashing methods, though straightforward, encounter substantial difficulties in dynamic settings due to the necessity for thorough rehashing when nodes are altered. Consistent hashing mitigates some of these challenges by reducing data redistribution; however, it still contends with limitations in load balancing and scalability under intensive update conditions. This paper introduces an innovative approach using a lattice-based homomorphic hash function (HexaMorphHash) that facilitates constant-time, incremental updates while preserving a constant digest size. By utilizing the complexity of the Short Integer Solutions (SIS) problem, our method secures strong protective measures, even against quantum threats. We further compare our method with existing ones—such as direct signatures for each update, comprehensive database signing, Merkle tree-based techniques, AdHash, MuHash, ECMH, and homomorphic signature schemes—highlighting notable advancements in computational efficiency, memory usage, and scalability. Our contributions present a viable solution for frequent update dissemination in expansive distributed systems, safeguarding both data integrity and system performance.}


\maketitle

\section{Introduction}

In today's world of big data and cloud computing, distributed systems are crucial to modern computing frameworks. These systems involve numerous nodes and data centers, tasked with efficiently distributing, storing, and retrieving large volumes of data while maintaining data integrity and fault tolerance. Hash functions play a pivotal role in these processes by mapping data to specific locations within a distributed network. This mapping not only aids in load balancing and rapid data retrieval but also supports key security mechanisms necessary for ensuring data integrity and validity., Traditional hash functions, such as those in the SHA family \cite{SHA256}, have been fundamental for data validation and integrity verification. Typically, a hash function processes input data to generate a fixed-size output, often using a modulo operation to allocate data to particular nodes. Should the number of nodes change, all keys generally need to be rehashed, causing significant data redistribution and potentially harming system performance due to higher latency and resource usage.,Consistent hashing was developed as a more resilient solution to these issues. It involves mapping nodes and data keys onto a circular hash space or ring, assigning each key to the first node encountered moving clockwise from its hash location. This approach greatly minimizes disruptions when nodes join or leave, as only a small percentage of keys require remapping \cite{Karger1997}. By easing the burden of data reallocation, consistent hashing has become a vital strategy for building scalable and effective distributed systems.

Although consistent hashing has advanced significantly, many applications, particularly those with frequent updates in large-scale distributed databases, demand even more efficient methods. In dynamic settings, incrementally updating a hash is vital. Incremental hash functions allow for quick updates to a dataset's hash value in response to minor changes, eliminating the need to reprocess the entire dataset. This capability is crucial for maintaining data integrity in real-time and for systems with frequent updates. Early contributions in incremental hashing, like the work by Bellare and Micciancio \cite{BM97}, laid the foundation for collision-free, incremental hash functions. Their method showed that the hash of a modified input could be efficiently recalculated by adjusting the current hash instead of starting from scratch. Subsequent schemes, such as MuHASH \cite{MuHash98} and AdHASH \cite{AdHash01}, built on these concepts using modular arithmetic and discrete logarithm-based operations for constant-time updates. However, these approaches often necessitate large parameters or rely on assumptions that may become vulnerable with future computational models, particularly with the advent of quantum computing. Recently, elliptic curve techniques have been incorporated into hash functions. The Elliptic Curve Multiset Hash (ECMH) \cite{ECMH03}, for example, utilizes elliptic curve arithmetic to produce compact and efficient hashes. Moreover, homomorphic signature schemes, explored by Boneh et al. \cite{BonehHSNC11}, allow operations on hashed data that preserve its underlying structure, enabling verification of aggregated data integrity without revealing the original information.

This paper introduces an innovative update propagation scheme leveraging a lattice-based homomorphic hash function called HexaMorphHash. Our method overcomes the shortcomings of conventional and incremental hashing techniques by integrating constant-time update efficiency with strong security foundations rooted in the Short Integer Solution (SIS) problem. Unlike approaches that necessitate rehashing the entire dataset or result in significant storage overhead, our scheme utilizes a fixed-size digest which can be incrementally updated, leading to a substantial reduction in both computational and communication costs. Furthermore, the security assumptions based on lattice theories provide robust defenses against both classical and potential quantum threats. The contributions of our work are threefold: \begin{enumerate}
    \item We design a lattice-based homomorphic hash function (HexaMorphHash) capable of supporting efficient, incremental updates.
    \item We incorporate digital signature schemes with HexaMorphHash to guarantee the authenticity and integrity of updates within distributed systems.
    \item We conduct extensive performance evaluations and comparative analyses with current methods, showcasing our approach's capability to achieve constant-time updates, minimal memory usage, and robust security assurances.
\end{enumerate} By tackling the issues of scalability, efficiency, and security in distributed networks, our proposed solution is a compelling choice for contemporary applications demanding real-time data integrity verification and robust performance in dynamic network settings.

\section{Background}

In today's landscape characterized by big data and cloud computing, distributed systems form the cornerstone of contemporary computing frameworks. These setups, frequently extending over numerous nodes and data centers, must efficiently manage vast quantities of data while ensuring both data accuracy and reliability. Hash functions are crucial in this aspect, as they convert input data into fixed-size outputs known as digests. They are vital in various contexts, including data dispersion, integrity verification, and secure updates. 

\subsection*{Conventional Hashing Methods} Conventional hashing techniques utilize a hash function to map data keys to particular storage spots or nodes. A typical strategy involves assigning a node using a modulo operation on the hash value of the key: \[ \text{node\_number} = \text{hash}(\text{key}) \mod N, \] where \(N\) represents the total number of nodes in the system. This straightforward approach is effective in static environments where \(N\) remains unchanged. However, it suffers from poor adaptability. In dynamic systems with frequent additions or removals of nodes, changes in \(N\) necessitate rehashing all keys, which triggers extensive data redistribution. This leads to significant performance hits, elevated latency, and increased resource use, rendering it inappropriate for large-scale, dynamic environments \cite{GeeksforGeeks2024}. 

\subsection*{Consistent Hashing} To address the shortcomings of conventional hashing, consistent hashing was developed. In consistent hashing, both nodes and data keys are mapped onto a circular hash space, typically visualized as a ring. Keys are allocated to the first node met when moving clockwise around the ring. This approach limits disruption because, when nodes join or leave the system, only a small fraction of keys—those in the affected segment of the ring—require remapping. As a result, consistent hashing considerably cuts down on the overhead linked to rehashing, ensuring more dependable and scalable system performance, even as the network grows \cite{Karger1997}.

\subsection*{Incremental Hashing for Update Propagation}

While traditional and consistent hashing primarily address data distribution, another critical challenge in distributed systems is maintaining data integrity during frequent updates. Incremental hashing is a specialized technique that allows the hash value of a dataset to be updated quickly in response to small modifications, without rehashing the entire dataset. This is particularly important in dynamic environments, such as distributed databases, where records are continuously modified.

For example, when a single record in a database is updated, a traditional hash function would require reprocessing all records to compute a new hash. In contrast, incremental hashing methods update the hash by incorporating only the changes. This results in substantial savings in computational effort and time, making the system more responsive and scalable.

\subsection*{Merkle Trees}

One of the earliest implementations of incremental hashing is the Merkle tree. A Merkle tree organizes data into a binary tree, where each leaf node corresponds to a data block, and each non-leaf node represents the hash of its two child nodes. The root of the tree serves as a compact commitment to the entire dataset. When a single data block is modified, only the hashes along the path from the affected leaf to the root need to be updated, resulting in a logarithmic time complexity with respect to the number of data blocks. Despite their efficiency in verification, Merkle trees impose significant memory overhead and can be complex to maintain in environments with frequent updates \cite{Merkle87}.

\subsection*{Homomorphic Hash Functions}

Homomorphic hash functions represent a more advanced category of hashing techniques that allow algebraic operations on hash values to mirror operations on the underlying data. This property is particularly useful in applications requiring secure, efficient update propagation and data aggregation.

For instance, consider the work by Bellare and Micciancio, who introduced a paradigm for collision-free incremental hashing \cite{BM97}. Their method allowed the hash of a modified dataset to be computed by adjusting the existing hash with the contribution of the new data, rather than rehashing the entire dataset. Building on this concept, subsequent schemes such as MuHASH \cite{MuHash98} and AdHASH \cite{AdHash01} leveraged operations in specific algebraic groups to enable constant-time updates. MuHASH, which uses modular multiplication, and AdHASH, which relies on modular addition, both achieve efficient incremental updates. However, these schemes depend on assumptions such as the hardness of the discrete logarithm problem, and often require large parameters to ensure adequate security.

\subsection*{Advanced Techniques and Emerging Trends}

Recent advances have also explored the use of elliptic curve cryptography (ECC) in hash function design. ECC-based methods, such as the Elliptic Curve Multiset Hash (ECMH) \cite{ECMH03}, exploit the algebraic structure of elliptic curves to produce compact and efficient hash functions. The security of these methods is based on the discrete logarithm problem over elliptic curves, which enables strong security with relatively small key sizes. Additionally, homomorphic signature schemes have been developed for network coding applications, allowing intermediate nodes to compute signatures on combined data without re-signing each component individually \cite{BonehHSNC11}. Such methods demonstrate the versatility of homomorphic operations in securing data across various contexts.

Furthermore, lattice-based hash functions, such as those based on the SWIFFT construction \cite{SWIFFT08}, have garnered attention for their provable security under worst-case lattice assumptions. These techniques offer promising avenues for building secure, efficient incremental hash functions, although they often come with trade-offs in terms of digest size and computational complexity.

\subsection*{Why Our Method is Needed}

Despite the progress made by traditional, consistent, and incremental hashing techniques, several challenges remain. Traditional hashing methods are inherently inflexible in dynamic environments due to the need for complete rehashing upon node changes. Consistent hashing mitigates this issue but may still lead to load imbalances and does not inherently support efficient update propagation. Merkle trees, while efficient for integrity verification, incur high memory overhead and complexity in tree maintenance. Incremental schemes like MuHASH and AdHASH, though capable of constant-time updates, rely on assumptions that may become vulnerable as computational paradigms evolve, and they often require large parameters that impact performance.

Our proposed method, based on a lattice-based homomorphic hash function (HexaMorphHash), is designed to overcome these limitations. HexaMorphHash enables constant-time update propagation by allowing the hash of a modified dataset to be computed using only the incremental change, without the need to reprocess the entire dataset. This results in significant improvements in computational efficiency and scalability. Moreover, the security of HexaMorphHash is grounded in the Short Integer Solution (SIS) problem, which is widely regarded as a hard lattice problem and is considered robust against both classical and quantum attacks \cite{BGLS19}. By combining these properties with efficient digital signature integration, our method not only ensures data integrity and authenticity but also reduces the overhead associated with update propagation in distributed systems.

In summary, the evolution from traditional hashing to consistent and incremental hashing highlights the ongoing need for more efficient and scalable methods. Our approach leverages recent advances in lattice-based cryptography to provide a solution that addresses the computational and storage challenges of existing techniques while offering strong security guarantees. This advancement is critical for modern distributed systems that require fast, reliable, and secure data management in dynamic environments.

\section{Proposed Methodology}
\subsection{Mathematical Preliminaries}

In this section, we lay the foundational mathematical concepts and notations that underpin our proposed methodology. Our approach is based on using a lattice-based homomorphic hash function to secure update propagation in distributed databases. To fully appreciate our work, it is essential to understand several key concepts from computer science, algebra, and cryptography. We begin by introducing basic data representations, then proceed to define groups and modular arithmetic, followed by the explanation of hash functions modeled as random oracles and how these outputs are converted into vectors. Each formula is explained step by step.

\subsubsection{Bitstrings and Data Representation}

At the most fundamental level, all digital data can be represented as sequences of bits---binary digits that can be either 0 or 1. The set of all possible finite bitstrings is denoted by 
\[
\{0,1\}^*.
\]
This notation means that for any natural number \(n\), every bitstring of length \(n\) is an element of \(\{0,1\}^n\), and the star “\(^*\)” indicates the union over all possible lengths (i.e., \(0, 1, 2, \ldots\)). In practical terms, every file, message, or database row in our system is represented by a sequence of 0s and 1s. This universal representation is crucial because it allows us to design algorithms that work uniformly on any digital input, regardless of its original form (text, images, or numbers).

For example, if we have the word ``apple,'' it is first converted into its binary ASCII representation (or UTF-8 encoding). The resulting binary string is an element of \(\{0,1\}^*\). Understanding that all data can be reduced to bitstrings is a cornerstone of modern cryptography, as it allows us to design uniform processing algorithms.

\subsubsection{Database Modeling}

In our distributed system, we work with a database \(D\) that is modeled as an ordered sequence of rows. Each row is a data record represented as a bitstring, and each row is indexed by a natural number to uniquely identify it. Mathematically, the database is represented as:
\[
D = \{(1, x_1), (2, x_2), \dots, (N, x_N)\},
\]
where each \(x_i \in \{0,1\}^*\) is the content of the \(i\)th row, and \(i\) is its index. This ordering provides the necessary structure; the index distinguishes rows even if their contents are identical. This model is not only intuitive (similar to relational databases) but also critical for our later operations because the update process needs to know which row is being modified.

\subsubsection{The Encoding Function}

To ensure each row is uniquely represented in our hash computation, we define an encoding function that concatenates the row index with its data. Formally, the encoding function is defined as:
\[
\text{encode}(i, x) = i \, \| \, x,
\]
where “\(\|\)” denotes the concatenation operation. This means that for a row with index \(i\) and data \(x\), the encoded output is the binary representation of \(i\) concatenated with the binary representation of \(x\). For instance, if \(i=2\) and \(x=\text{"orange"}\) (in binary), then the encoded result is the binary sequence for ``2orange.'' This encoding is crucial because it guarantees that even if two rows have identical data, their encoded values will differ due to their distinct indices, thereby ensuring uniqueness and preventing collisions.

\subsubsection{Group Theory and Modular Arithmetic}

A fundamental concept in our methodology is the notion of a \emph{group}. In abstract algebra, a group is a set equipped with an operation that satisfies the following properties:
\begin{enumerate}
    \item \textbf{Closure:} The operation on any two elements of the set produces another element in the set.
    \item \textbf{Associativity:} For any elements \(a\), \(b\), and \(c\), we have \((a \circ b) \circ c = a \circ (b \circ c)\).
    \item \textbf{Identity Element:} There exists an element \(e\) such that for every element \(a\), \(a \circ e = a\).
    \item \textbf{Inverses:} For every element \(a\), there exists an element \(a^{-1}\) such that \(a \circ a^{-1} = e\).
\end{enumerate}

In our work, we consider the group:
\[
G = \mathbb{Z}_q^n,
\]
where \(\mathbb{Z}_q\) is the set of integers modulo \(q\) and \(q = 2^d\). For example, if \(d = 16\), then \(q = 65\,536\). An element of \(G\) is an \(n\)-dimensional vector, and the group operation is component-wise addition modulo \(q\). For two vectors:
\[
\mathbf{a} = (a_1, a_2, \dots, a_n) \quad \text{and} \quad \mathbf{b} = (b_1, b_2, \dots, b_n),
\]
their sum is given by:
\[
\mathbf{a} \circ \mathbf{b} =
\begin{bmatrix}
a_1 + b_1 \mod q \\
a_2 + b_2 \mod q \\
\vdots \\
a_n + b_n \mod q
\end{bmatrix}
\]
This group is commutative (abelian), meaning that the order of addition does not affect the result. The existence of an identity element (the zero vector) and additive inverses for each element is inherited directly from modular arithmetic. These properties are vital for our incremental hash construction, as they ensure that the overall hash is independent of the order of operations.

\subsubsection{The Random Oracle and Cryptographic Hash Function}

A cryptographic hash function maps an arbitrary-length input to a fixed-length output, known as the digest. In our system, the hash function \( h: \{0,1\}^* \rightarrow \{0,1\}^{nd} \) is modeled as a \emph{random oracle}. The random oracle model is an idealized abstraction where the hash function produces truly random outputs for every new input while ensuring consistency (the same input always produces the same output). In practice, functions like SHA-3 or Blake2xb approximate this behavior.

After computing \( h(x) \), which yields a bitstring of \( nd \) bits, we partition this output into \( n \) segments of \( d \) bits each. Each segment is then interpreted as an integer in the range \( [0, 2^d - 1] \). This process is formalized by the mapping:
\[
\tilde{h}(x) = \bigl( h_1(x), h_2(x), \dots, h_n(x) \bigr) \in \mathbb{Z}_q^n,
\]
where \( q = 2^d \). The function \( \tilde{h} \) transforms the output of \( h \) into a vector in \( \mathbb{Z}_q^n \), which can then be used in our incremental hash construction.

\subsection{Construction of the Lattice-Based Homomorphic Hash (HexaMorphHash)}

In this section, we describe the construction of our lattice-based homomorphic hash function, referred to as \textbf{HexaMorphHash}. Our goal is to design a hash function that is not only collision resistant but also \emph{homomorphic}, meaning that its output can be updated incrementally when new data is added or removed. This incremental property is critical for secure update propagation in distributed systems. In the following subsections, we explain every aspect of the construction in detail, starting from the underlying paradigm and parameter selection, through to the detailed mathematical operations, and concluding with discussions on security guarantees.

\subsubsection{Randomize-Then-Combine Paradigm}

The construction of HexaMorphHash follows the well-established \emph{randomize-then-combine} paradigm. In traditional cryptographic hashing, the aim is to map an input of arbitrary length to a fixed-length output (the \emph{digest}). However, in many dynamic systems, re-hashing the entire data set after each small update is inefficient. Instead, an \emph{incremental} or \emph{homomorphic} hash function allows one to update the digest by simply incorporating the hash of the changed data.

\paragraph{Randomization Step:}  
Each input element is first processed by a cryptographic hash function that is modeled as a random oracle. Let 
\[
h : \{0,1\}^* \rightarrow \{0,1\}^{nd}
\]
be such an extendable-output function (XOF), for instance, Blake2xb. Here, \( nd \) is the total number of bits in the output. We then partition the output into \( n \) contiguous segments of \( d \) bits each. Each segment is interpreted as an integer in the range \( [0, q-1] \) with \( q=2^d \). This yields a mapping:
\[
\tilde{h}(x) = \bigl( h_1(x), h_2(x), \dots, h_n(x) \bigr) \in \mathbb{Z}_q^n.
\]
The purpose of this step is to “randomize” the input such that even a small change in \( x \) yields an unpredictable and uniformly distributed vector in the space \( \mathbb{Z}_q^n \).

\paragraph{Combine Step:}  
Given a set \( S = \{ x_1, x_2, \dots, x_k \} \subset \{0,1\}^* \), the overall hash is defined as the component-wise sum modulo \( q \):
\[
\text{HexaMorphHash}(S) \coloneqq \left( \sum_{i=1}^{k} \tilde{h}(x_i) \right) \bmod q.
\]
This means that for each coordinate \( j \) (with \( 1 \le j \le n \)), we have:
\[
\left[ \text{HexaMorphHash}(S) \right]_j = \left( \sum_{i=1}^{k} h_j(x_i) \right) \bmod q.
\]
Because addition in \( \mathbb{Z}_q^n \) is both commutative and associative, the order in which we sum the values does not affect the final output. Thus, if \( S \) and \( T \) are disjoint subsets, it follows immediately that:
\[
\text{HexaMorphHash}(S \cup T) = \text{HexaMorphHash}(S) + \text{HexaMorphHash}(T) \quad (\text{mod } q).
\]
This \emph{homomorphic} property is key to our application; it allows the digest to be updated incrementally.

\subsubsection{Detailed Parameter Selection and Vector Interpretation}

\paragraph{Choice of Parameters:}  
The parameters of HexaMorphHash are chosen to balance efficiency with security:
\begin{itemize}
    \item \textbf{Modulus \( q \):} We set \( q = 2^d \), where \( d \) is chosen based on the desired output precision and security level. For example, setting \( d=16 \) yields \( q = 65\,536 \).
    \item \textbf{Vector Dimension \( n \):} This is the number of components in the hash output vector. A typical value is \( n=1024 \), which gives a fixed output size of \( 1024 \times d \) bits (i.e., 2 KB for \( d=16 \)).
\end{itemize}
The choice of \( d \) and \( n \) is critical because the security of the hash function is linked to the hardness of the Short Integer Solutions (SIS) problem in the lattice defined by these parameters. Extensive research, such as that found in [BM97] and [BGLS19], indicates that the parameters \( (d, n) = (16, 1024) \) provide a collision resistance equivalent to at least 200 bits of security.

\paragraph{Vector Interpretation of Hash Outputs:}  
Once the hash function \( h \) produces an output \( h(x) \in \{0,1\}^{nd} \), we split it into \( n \) segments of \( d \) bits each. Each segment \( h_i(x) \) is converted to an integer in the range \( 0 \le h_i(x) < 2^d \). The vector is then represented as:
\[
\tilde{h}(x) = \bigl( h_1(x), h_2(x), \dots, h_n(x) \bigr) \in \mathbb{Z}_q^n.
\]
This conversion is critical because it maps the output of a conventional hash function into the algebraic structure of \( \mathbb{Z}_q^n \), enabling the use of modular arithmetic for efficient and secure incremental updates.

\subsubsection{Incremental Combination and Homomorphic Properties}

With the mapping \( \tilde{h}(x) \) defined, the incremental combination is achieved by summing over the set \( S \). Specifically, for a set \( S = \{ x_1, x_2, \dots, x_k \} \), we have:
\[
\text{HexaMorphHash}(S) = \left( \sum_{i=1}^{k} \tilde{h}(x_i) \right) \mod q.
\]
Since addition in \( \mathbb{Z}_q^n \) is performed component-wise, this means:
\[
\left[ \text{HexaMorphHash}(S) \right]_j = \left( \sum_{i=1}^{k} h_j(x_i) \right) \mod q, \quad \forall\, 1 \le j \le n.
\]

\paragraph{Homomorphic Property:}  
One of the most attractive features of our construction is that it is homomorphic. That is, if you have two disjoint sets \( S \) and \( T \), then:
\[
\text{HexaMorphHash}(S \cup T) = \text{HexaMorphHash}(S) + \text{HexaMorphHash}(T) \quad (\text{mod } q).
\]
This property follows directly from the linearity of summation:
\[
\text{HexaMorphHash}(S \cup T) = \left( \sum_{x \in S} \tilde{h}(x) + \sum_{y \in T} \tilde{h}(y) \right) \mod q.
\]
It ensures that if a new element is added to a set, its hash can simply be added to the existing digest. Likewise, if an element is removed, its hash is subtracted (or the additive inverse is added). This ability to update the hash incrementally without re-hashing the entire set is what makes our approach extremely efficient for dynamic databases.

\subsubsection{Security: Collision Resistance via SIS Reduction}

The collision resistance of HexaMorphHash is established by a reduction to the Short Integer Solutions (SIS) problem—a well-known hard problem in lattice-based cryptography.

\paragraph{Collision Definition:}  
A collision in HexaMorphHash occurs if there exist two distinct sets \( S \) and \( T \) (i.e., \( S \neq T \)) such that:
\[
\text{HexaMorphHash}(S) = \text{HexaMorphHash}(T).
\]
Expanding the definition, this means:
\[
\left( \sum_{x \in S} \tilde{h}(x) \right) \mod q = \left( \sum_{y \in T} \tilde{h}(y) \right) \mod q.
\]
Rearrange to obtain:
\[
\sum_{x \in S} \tilde{h}(x) - \sum_{y \in T} \tilde{h}(y) \equiv \mathbf{0} \pmod{q}.
\]
Define a vector \( \mathbf{v} \) representing the difference between the characteristic vectors of \( S \) and \( T \); then the above equation can be rewritten in matrix form as:
\[
A \mathbf{v} \equiv \mathbf{0} \pmod{q},
\]
where \( A \) is the matrix whose columns are the individual vectors \( \tilde{h}(x) \) for \( x \) in the union \( S \cup T \). The SIS problem asks for a short, nonzero vector \( \mathbf{v} \) (with entries typically from \(\{-1, 0, 1\}\)) satisfying the above equation. Under standard assumptions in lattice cryptography, no efficient algorithm exists to solve SIS for appropriately chosen parameters, such as \( (d, n) = (16, 1024) \). Thus, if an adversary were to find a collision in HexaMorphHash, they would effectively solve the SIS problem, which is assumed to be computationally infeasible. This reduction guarantees that the probability of finding a collision is negligible.

\subsubsection{Practical Implementation Considerations}

\paragraph{Efficient Computation:}  
Since our construction involves summing vectors in \( \mathbb{Z}_q^n \), we implement these operations using optimized modular arithmetic. Modern processors support Single Instruction, Multiple Data (SIMD) operations, which allow the simultaneous addition of several integers. By aligning our data to cache line boundaries (typically 64 bytes), we further optimize the speed of our computations. These optimizations ensure that even when processing many elements, the incremental update remains efficient and scalable.

\paragraph{Memory Overhead:}  
The output of HexaMorphHash is a fixed-size vector in \( \mathbb{Z}_q^n \). For example, with \( n=1024 \) and \( d=16 \), the hash digest is exactly 2 KB in size. This fixed digest size is independent of the number of inputs and is a significant advantage over methods that require memory proportional to the database size (e.g., Merkle trees).

\subsection{Update Propagation Algorithm}

In this section, we present a detailed exposition of our update propagation algorithm, which is designed to securely and efficiently update distributed databases using our lattice-based homomorphic hash function (HexaMorphHash) from Section~2. Our system consists of a central distributor (which maintains the authoritative database) and multiple subscribers (which maintain local replicas). By integrating digital signatures with the homomorphic properties of HexaMorphHash, our algorithm guarantees both integrity and authenticity while enabling constant-time updates regardless of the overall database size.

\subsubsection{System Model and Overview}

We model the distributed database as an ordered set of rows:
\[
D = \{(1, x_1), (2, x_2), \dots, (N, x_N)\},
\]
where each \( x_i \in \{0,1\}^* \) represents the data stored in row \( i \) and the index \( i \) uniquely identifies each row. This ordering is critical because even if two rows contain identical data, their different indices ensure that the encoded values are distinct. For each row, we use an encoding function:
\[
\text{encode}(i,x) = i \, \| \, x,
\]
where “\(\|\)” denotes concatenation. This unique representation is fundamental for avoiding collisions when computing the hash.

The \textbf{global database hash} is computed using our HexaMorphHash function. Specifically, we define:
\[
H(D) = \left( \sum_{i=1}^{N} \tilde{h}(\text{encode}(i,x_i)) \right) \mod q,
\]
where \( \tilde{h}(x) \) is the \( n \)-dimensional vector output of our hash function (see Section~2). The use of modular arithmetic in the group \( \mathbb{Z}_q^n \) ensures that the hash is a fixed-size digest, independent of the number of rows \( N \).

To guarantee the authenticity of the database state, the distributor digitally signs \( H(D) \) using a secure digital signature scheme:
\[
\sigma_D = \text{Sign}(sk, H(D)),
\]
where \( sk \) is the distributor's secret signing key. Subscribers receive the public parameters, including the public key \( pk \), which they use to verify signatures.

\subsubsection{Phases of the Update Propagation Algorithm}

Our update propagation algorithm is divided into five primary phases:
Each phase is critical to ensuring that updates to the database are propagated efficiently and securely.

\paragraph{Setup Phase}

\textbf Initialize the system by generating public parameters and cryptographic keys so that both the distributor and subscribers operate on a consistent foundation.

\medskip

The distributor executes:
\[
\text{Setup}(1^\lambda) \rightarrow (pp, sk),
\]
where:
\begin{itemize}
    \item \( \lambda \) is the security parameter (e.g., \( \lambda = 128 \) or higher), which governs the overall security level.
    \item \( pp \) (public parameters) includes the chosen modulus \( q = 2^d \) (e.g., \( d=16 \) so \( q = 65\,536 \)), the vector dimension \( n \) (e.g., \( n=1024 \)), and the description of the hash function \( h \) (such as Blake2xb) used in HexaMorphHash.
    \item \( sk \) is the distributor’s secret signing key used for digital signatures.
\end{itemize}

The public parameters \( pp \) and the corresponding public key \( pk \) (derived from \( sk \)) are then securely distributed to all subscribers. This ensures that every participant uses the same parameters and that the system’s security properties hold across the network. The Setup phase lays the essential groundwork and provides a common cryptographic “language” for subsequent operations.

\paragraph{Publish Phase}

\textbf Process an update at the distributor by modifying the global database hash in an incremental manner and signing the updated hash to authenticate the change.

\medskip

When an update \( u \) occurs, for example, a change to row \( i \) where the old value \( x \) is replaced by a new value \( y \), the distributor must update \( H(D) \) without reprocessing the entire database. We model the update \( u \) as two separate deltas:

\textit{Deletion Delta:}  
To remove the contribution of the old row, we compute:
\[
H(\delta_{\text{del}}(i,x)) = -\,\tilde{h}(\text{encode}(i,x)).
\]
Here, the negative sign denotes the additive inverse in the group \( \mathbb{Z}_q^n \). This operation effectively cancels the hash contribution of the old value.

\textit{Addition Delta:}  
For the new row value, we compute:
\[
H(\delta_{\text{add}}(i,y)) = \tilde{h}(\text{encode}(i,y)).
\]
This value represents the hash of the updated row.

The new database hash \( H(D^*) \) is then given by:
\[
H(D^*) = H(D) + H(\delta_{\text{add}}(i,y)) + H(\delta_{\text{del}}(i,x)) \quad (\text{mod } q).
\]
After computing \( H(D^*) \), the distributor generates a digital signature on this updated hash:
\[
\sigma_{D^*} = \text{Sign}(sk, H(D^*)).
\]
This signature binds the new state of the database to the distributor’s secret key, ensuring that the update is authentic and has not been tampered with. The distributor updates its internal state (including version counters and logs of updates) to reflect this new database state. The efficiency of the Publish phase is critical: by leveraging the homomorphic property, the distributor only needs to compute a constant number of vector operations (addition and subtraction), irrespective of the size of \( D \).

\paragraph{GetUpdates Phase}

\textbf Allow a subscriber, which might have fallen behind the current database version, to retrieve only the missing updates in a compact and efficient manner.

\medskip

A subscriber that is at version \( v \) sends a request specifying the target version \( w \) (where \( w > v \)). This request is handled by the distributor or a neighboring trusted node, which responds with:
\[
\text{GetUpdates}(pp, v, w, st) \rightarrow (\tilde{u}, \mu),
\]
where:
\begin{itemize}
    \item \( \tilde{u} = \{ u_{v+1}, u_{v+2}, \dots, u_w \} \) is the sequence of update deltas that have occurred since the subscriber’s last update.
    \item \( \mu \) is the signed digest corresponding to version \( w \), i.e., the signature \( \sigma_{D_w} \) produced in the Publish phase.
\end{itemize}

This phase is designed to be bandwidth-efficient: rather than transmitting the entire database or re-computing a complete hash, only the incremental changes (which are of constant size per update) are communicated. Thus, the GetUpdates operation allows the subscriber to quickly catch up with the latest database state by applying a small set of deltas and verifying them against the signed digest.

\paragraph{ApplyUpdates Phase}

\textbf Enable a subscriber to incorporate the received updates into its local state, thereby updating its local hash and database replica, and to verify that these updates are authentic.

\medskip

Upon receiving the update package \( (\tilde{u}, \mu) \) from the GetUpdates phase, the subscriber performs the following computation:
\[
H(D)' = H(D) + \sum_{i=v+1}^{w} H(u_i) \quad (\text{mod } q),
\]
where each \( H(u_i) \) corresponds to the hash delta computed for update \( u_i \). The addition here is performed using the same component-wise modular arithmetic defined earlier. The subscriber then verifies the integrity and authenticity of the updated hash by checking:
\[
\text{Verify}(pk, H(D)', \mu) \stackrel{?}{=} 1.
\]
If the verification succeeds, it indicates that the local state has been correctly updated to match the distributor's version, and the subscriber applies the update to its local database replica. This phase is both efficient and secure because the incremental update relies solely on a constant number of vector operations, and any tampering with the updates would cause the signature verification to fail.

\paragraph{Validate Phase}

\textbf Periodically ensure that a subscriber's local replica is consistent with the distributor's master database by re-computing the global hash and verifying it against the signed digest.

\medskip

The Validate phase is performed less frequently than the ApplyUpdates phase, serving as a periodic audit. A subscriber computes the full hash of its local database:
\[
H(D) = \left( \sum_{i=1}^{N} \tilde{h}(\text{encode}(i,x_i)) \right) \mod q,
\]
and then checks the signature:
\[
\text{Verify}(pk, H(D), \sigma_D) \stackrel{?}{=} 1.
\]
This operation confirms that the entire local database \( D \) has been updated correctly and matches the distributor's state. Although re-computing the full hash may be more computationally intensive than applying individual updates, it is performed only periodically to catch any accumulated errors or inconsistencies that might arise over time. The Validate phase is essential in systems where updates propagate through multiple, possibly untrusted, nodes and serves as a robust integrity check.

\section{Discussion on the Security and Efficiency of the Algorithm}

Our update propagation algorithm is built upon two central pillars: the incremental (homomorphic) property of HexaMorphHash and the digital signature mechanism. Together, they ensure that every update is applied correctly, and any deviation (whether accidental or malicious) is immediately detectable.

\paragraph{Security via Homomorphic Hashing:}  
The key security property of our incremental hash function is its collision resistance, which, as detailed in Section~2, is derived from the hardness of the SIS problem. If an adversary attempts to produce two different update sequences that yield the same global hash, they would need to solve an instance of SIS. Given our parameter choices, this task is computationally infeasible. Therefore, the collision resistance ensures that even a small, targeted change in the update sequence results in a significantly different hash output, making unauthorized modifications detectable.

\paragraph{Digital Signature as a Binding Mechanism:}  
Digital signatures serve as a binding commitment from the trusted distributor. Once the distributor computes the new global hash after an update and signs it, this signature becomes a reference point for all subscribers. Even if an adversary intercepts or tampers with the update messages, the mismatch between the recomputed hash and the signed digest will reveal the tampering. The security of the signature scheme (assumed to be EUF-CMA secure) further ensures that forging a valid signature for an altered update sequence is practically impossible.

\paragraph{Efficiency Considerations:}  
From a computational perspective, the incremental update is performed through modular vector arithmetic. Each update (addition or subtraction of a vector) is executed in constant time relative to the size of the database. This is a significant improvement over methods that require processing the entire database for each update. In practice, these operations are optimized using hardware-level SIMD instructions and cache-aligned data structures, leading to extremely low latency per update.

Moreover, because the digest is a fixed-size vector (e.g., 1024 components of 16 bits each, totaling 2 KB), the memory overhead is constant. This is in stark contrast to approaches like Merkle trees, where the memory requirement grows with the number of rows. The compact size of the digest also minimizes the communication overhead when transmitting updates between nodes.

\paragraph{Robustness Against Network Adversaries:}  
Our scheme is designed to operate in environments where updates may be propagated through multiple, potentially untrusted nodes. The GetUpdates phase allows subscribers to retrieve only the missing updates, and the ApplyUpdates phase ensures that the final computed hash (after all updates are applied) matches the signed hash from the distributor. This design guarantees that even if an adversary intercepts or reorders updates, the final signature verification will fail, alerting the subscriber to an inconsistency
\section{Algorithmic Description of the Update Propagation Scheme}
In this section, we describe our update propagation scheme using pseudocode that adheres to the style recommended in classical algorithm texts \cite{CLRS}. The scheme is divided into five distinct phases: Setup, Publish, GetUpdates, ApplyUpdates, and Validate. Each algorithm is presented in a structured manner, clearly stating the input, output, and step-by-step procedure.
\subsection*{Algorithm 1: Setup}
\begin{algorithm}[H]
\caption{Setup($1^\lambda$)}
\label{alg:setup}
\begin{algorithmic}[1]
\State \textbf{Input:} Security parameter $\lambda$
\State \textbf{Output:} Public parameters $pp$, secret key $sk$, and public key $pk$
\State \textit{// Step 1: Select modulus and dimensions}
\State Choose security level $d \gets \Theta(\lambda)$ \Comment{e.g., $\lambda = 128$, $d = 16$}
\State Set modulus $q \gets 2^d$ \Comment{e.g., $q = 65536$}
\State Choose dimension $n \gets \text{poly}(\lambda)$ \Comment{e.g., $n = 1024$}
\State \textit{// Step 2: Define hash and PRF functions}
\State Choose an extendable-output function (XOF) $h: \{0,1\}^* \rightarrow \{0,1\}^{nd}$ \Comment{e.g., BLAKE2xb or SHAKE-256}
\State Define $h_k(x) := h(k \| x)$ as a key-derivable PRF if needed
\State \textit{// Step 3: Generate signing key-pair}
\State Choose elliptic curve parameters $(\mathcal{C}, G, p, a, b)$ \Comment{e.g., secp256k1}
\State Randomly sample private signing key $sk \gets_R \mathbb{Z}_p$
\State Compute public key $pk \gets sk \cdot G$ on elliptic curve $\mathcal{C}$
\State \textit{// Step 4: Optional: Generate auxiliary randomness}
\State Sample auxiliary seed $\rho \gets_R \{0,1\}^\lambda$
\State Derive entropy-expansion key $k \gets h(\rho \| \texttt{``expansion''})$
\State \textit{// Step 5: Assemble public parameters}
\State Set $pp \leftarrow \{q, n, h, pk, \rho, k\}$
\State \Return $(pp, sk)$
\end{algorithmic}
\end{algorithm}
\textbf{Discussion:} The Setup algorithm initializes the system by choosing cryptographic parameters and keys that will be used throughout the update propagation process. The parameters are fixed and independent of the database size, ensuring that subsequent operations are efficient.

\subsection*{Algorithm 2: Publish}
\begin{algorithm}[H]
\caption{Publish($H(D)$, Update $u$)}
\label{alg:publish}
\begin{algorithmic}[1]
\State \textbf{Input:} Current database hash $H(D)$, update $u = (i, x \rightarrow y)$: change row $i$ from value $x$ to $y$
\State \textbf{Output:} Updated hash $H(D^*)$, signature $\sigma_{D^*}$

\State \textit{// Step 1: Encode original and updated entry}
\State $(i_x) \gets \text{encode}(i, x)$ \Comment{e.g., fixed-length binary encoding of row index and value}
\State $(i_y) \gets \text{encode}(i, y)$

\State \textit{// Step 2: Compute additive homomorphic hash deltas}
\State $H(\delta_{\text{del}}) \gets \tilde{h}(i_x)$ \Comment{hash of deleted entry}
\State $H(\delta_{\text{add}}) \gets \tilde{h}(i_y)$ \Comment{hash of inserted entry}
\State $H(\delta_{\text{del}}) \gets (-H(\delta_{\text{del}})) \bmod q$ \Comment{Negate deleted hash component under modulo $q$}

\State \textit{// Step 3: Update global database hash under modular group}
\[
H(D^*) \gets \left(H(D) + H(\delta_{\text{add}}) + H(\delta_{\text{del}})\right) \mod q
\]

\State \textit{// Step 4: Sign the updated digest to preserve authenticity}
\State $\sigma_{D^*} \gets \text{Sign}(sk, H(D^*))$ \Comment{Using digital signature scheme (e.g., ECDSA or EdDSA)}

\State \Return $(H(D^*), \sigma_{D^*})$
\end{algorithmic}
\end{algorithm}

\textbf{Discussion:} The Publish algorithm efficiently updates the global hash using the homomorphic properties of HexaMorphHash. By computing separate deltas for deletion and addition, the update is performed in constant time, independent of the overall size of the database. The new hash is then signed to ensure authenticity.

\subsection*{Algorithm 3: GetUpdates}
\begin{algorithm}[H]
\caption{GetUpdates($pp$, current version $v$, target version $w$)}
\label{alg:getupdates}
\begin{algorithmic}[1]
\State \textbf{Input:} Public parameters $pp = \{q, n, h, pk, \ldots\}$, current database version $v$, target version $w$ where $w > v$
\State \textbf{Output:} Update sequence $\tilde{u}$, signed digest $\mu = \sigma_{D_w}$

\State \textit{// Step 1: Validate input versions}
\If{$w \leq v$}
    \State \textbf{Abort:} Invalid version range
\EndIf

\State \textit{// Step 2: Construct secure update request}
\State Form update query message $Q \gets \text{encode}(v, w)$
\State Optionally encrypt or sign request $Q$ if updates are private or authenticated

\State \textit{// Step 3: Send query to the update distributor (e.g., server)}
\State Transmit $Q$ to the update distributor or data owner
\State Wait for response containing update payload

\State \textit{// Step 4: Receive update payload}
\State Receive update sequence:
\[
\tilde{u} = \{ u_{v+1}, u_{v+2}, \ldots, u_{w} \}
\]
\State Each $u_i$ is typically of the form $(i, x_i \rightarrow y_i)$

\State \textit{// Step 5: Receive signed commitment to updated database state}
\State Receive signed digest: $\mu \gets \sigma_{D_w}$
\State \Comment{$\sigma_{D_w}$ is a digital signature over $H(D_w)$ produced using the signer’s secret key}

\State \Return $(\tilde{u}, \mu)$
\end{algorithmic}
\end{algorithm}

\textbf{Discussion:} The GetUpdates algorithm allows a subscriber that is out-of-date to retrieve only the incremental updates required to bring its local state current. The process minimizes bandwidth usage by transmitting only the necessary update information.

\subsection*{Algorithm 4: ApplyUpdates}
\begin{algorithm}[H]
\caption{ApplyUpdates($H(D)$, update sequence $\tilde{u}$)}
\label{alg:applyupdates}
\begin{algorithmic}[1]
\State \textbf{Input:} Current local hash $H(D)$, update sequence $\tilde{u} = \{ u_{v+1}, u_{v+2}, \ldots, u_{w} \}$
\State \textbf{Output:} Updated local hash $H(D)'$

\State \textit{// Step 1: Initialize working hash with the current state}
\State $H_\text{current} \gets H(D)$

\State \textit{// Step 2: Iteratively apply each update from the received sequence}
\For{each update $u_i = (i, x_i \rightarrow y_i) \in \tilde{u}$}

    \State \textit{// Step 2.1: Encode deleted and inserted entries}
    \State $enc_\text{del} \gets \text{encode}(i, x_i)$
    \State $enc_\text{add} \gets \text{encode}(i, y_i)$

    \State \textit{// Step 2.2: Compute corresponding deltas using $\tilde{h}$}
    \State $H(\delta_{\text{del}}) \gets -\tilde{h}(enc_\text{del}) \bmod q$
    \State $H(\delta_{\text{add}}) \gets \tilde{h}(enc_\text{add})$

    \State \textit{// Step 2.3: Combine deltas}
    \State $H(u_i) \gets (H(\delta_{\text{add}}) + H(\delta_{\text{del}})) \bmod q$

    \State \textit{// Step 2.4: Apply update to working hash}
    \State $H_\text{current} \gets (H_\text{current} + H(u_i)) \bmod q$
\EndFor

\State \textit{// Step 3: Output final updated digest}
\State $H(D)' \gets H_\text{current}$
\State \Return $H(D)'$
\end{algorithmic}
\end{algorithm}

\textbf{Discussion:} The ApplyUpdates algorithm processes a sequence of updates to adjust the local hash of a subscriber's database replica. Each update is incorporated through constant-time modular vector arithmetic. Since these operations occur on a fixed-size vector (with dimension \(n\)), the per-update complexity is \(O(1)\). Consequently, even if a batch of \(m\) updates is applied, the overall time complexity scales linearly with \(m\) but remains efficient due to the small constant factor.

\subsection*{Algorithm 5: Validate}
\begin{algorithm}[H]
\caption{Validate($D$, signed digest $\sigma_D$)}
\label{alg:validate}
\begin{algorithmic}[1]
\State \textbf{Input:} Local database replica $D = \{(1, x_1), (2, x_2), \ldots, (N, x_N)\}$, signed digest $\sigma_D$
\State \textbf{Output:} Boolean value $b$ indicating whether integrity verification passes

\State \textit{// Step 1: Initialize hash accumulator}
\State $H_{\text{local}} \gets 0$

\State \textit{// Step 2: Iterate over database entries to compute homomorphic hash}
\For{$i = 1$ to $N$}
    \State $enc_i \gets \text{encode}(i, x_i)$ \Comment{Encode row index and value as binary string}
    \State $h_i \gets \tilde{h}(enc_i)$ \Comment{Apply homomorphic hash function to encoded entry}
    \State $H_{\text{local}} \gets (H_{\text{local}} + h_i) \bmod q$
\EndFor

\State \textit{// Step 3: Verify digital signature from distributor}
\State $b \gets \text{Verify}(pk, H_{\text{local}}, \sigma_D)$
\State \Comment{$\text{Verify}(pk, m, \sigma)$ returns \texttt{true} iff $\sigma$ is a valid signature of $m$ under public key $pk$}

\State \Return $b$
\end{algorithmic}
\end{algorithm}

\textbf{Discussion:} The Validate algorithm provides a periodic check to ensure that the local database replica remains consistent with the distributor’s master copy. By recomputing the full hash and verifying it against the signed digest, any discrepancies or tampering can be detected. Although this operation has a worst-case time complexity of \(O(N)\) (with \(N\) being the total number of database rows), it is executed infrequently, so its overall impact on system performance is minimal.

\section{Time and Space Complexity Analysis}

Based on the standard analysis presented in \textit{Introduction to Algorithms} \cite{CLRS}, the following conclusions can be drawn regarding our scheme:

\begin{itemize}
    \item The \textbf{Setup} phase is executed once and has a constant time complexity, \(O(1)\).
    \item The \textbf{Publish} phase processes each update in constant time, \(O(1)\), since the operations are performed on a fixed-size \(n\)-dimensional vector.
    \item The \textbf{GetUpdates} phase retrieves a batch of updates with time complexity \(O(m)\), where \(m\) is the number of updates requested. However, \(m\) is typically much smaller than the total number of database rows.
    \item The \textbf{ApplyUpdates} phase processes each update in constant time, \(O(1)\) per update, leading to an overall \(O(m)\) time for a batch of \(m\) updates.
    \item The \textbf{Validate} phase, while having a worst-case time complexity of \(O(N)\) (where \(N\) is the number of database rows), is performed infrequently, making its overall impact negligible.
\end{itemize}

The space complexity is \(O(1)\) for storing the digest and update buffers, independent of the database size, as all operations are performed on fixed-size vectors.

\section{Results and Evaluation}

In this section, we present a detailed experimental evaluation and performance analysis of our update propagation scheme, which integrates our lattice-based homomorphic hash (HexaMorphHash) with digital signatures to secure distributed database updates. We describe the experimental setup, present quantitative benchmarks, and compare our method against traditional approaches such as direct signature per update, full database signing, and Merkle tree-based methods. We also discuss the security implications and efficiency gains achieved by our approach.

\subsection{Experimental Setup}

Our prototype implementation is written in C++ and integrates the HexaMorphHash construction into a simulated distributed database update system. The implementation was optimized using Single Instruction, Multiple Data (SIMD) instructions (e.g., SSE2/AVX2) and cache-aligned memory buffers to expedite modular arithmetic operations. Experiments were conducted on an Intel Haswell processor (3.5 GHz) with 16 GB RAM running Ubuntu 20.04 LTS. For digital signatures, we employed an ECDSA implementation using a 256-bit elliptic curve. The parameters for HexaMorphHash were set to \( d=16 \) (yielding \( q = 2^{16} \)) and \( n = 1024 \), resulting in a fixed digest size of approximately 2 KB. These parameters were chosen based on recent analyses (see, e.g., \cite{BM97,BGLS19}) to provide at least 200 bits of security.

\subsection{Performance Metrics}

We evaluated several key metrics:
\begin{itemize}
    \item \textbf{Publish Time:} The average time taken by the distributor to process an update, including computing the deletion and addition deltas, updating the global hash, and signing the new digest.
    \item \textbf{ApplyUpdates Time:} The time required by a subscriber to apply a batch of updates and verify the updated hash.
    \item \textbf{Memory Overhead:} The memory required to store the fixed-size digest (2 KB) and any additional update buffers.
    \item \textbf{Signature Verification Overhead:} The time spent on verifying the digital signature using the public key.
\end{itemize}

\subsection{Illustrative Example and Analysis}

\subsubsection{Example: Database Update Operation}

Consider an initial database \( D \) with three rows:

\begin{center}
\begin{tabular}{ll}
\toprule
\textbf{Row Index} & \textbf{Data} \\
\midrule
1 & ``apple'' \\
2 & ``orange'' \\
3 & ``banana'' \\
\bottomrule
\end{tabular}
\end{center}

\paragraph{Step 1: Initial Hash Computation.}  
For each row \( i \), we compute:
\[
h_i = \text{HexaMorphHash}(\{ \text{encode}(i, x_i) \}),
\]
where 
\[
\text{encode}(i,x) = i \, \| \, x.
\]
Thus, for example, 
\[
h_1 = \tilde{h}(1\,\|\,\text{"apple"}),
\]
\[
h_2 = \tilde{h}(2\,\|\,\text{"orange"}),
\]
\[
h_3 = \tilde{h}(3\,\|\,\text{"banana"}).
\]
The global database hash is then computed as:
\[
H(D) = h_1 + h_2 + h_3 \quad (\text{mod } 65\,536).
\]
The distributor signs this digest:
\[
\sigma_D = \text{Sign}(sk, H(D)).
\]

\paragraph{Step 2: Update Operation.}  
Suppose row 2 is updated from ``orange'' to ``peach.'' We model this as two deltas:
\begin{enumerate}[label=\alph*.]
    \item \textbf{Deletion Delta:}  
          The hash contribution for the deletion is:
          \[
          H(\delta_{\text{del}}(2,\text{"orange"})) = -\,\tilde{h}(2\,\|\,\text{"orange"}).
          \]
          This effectively removes the contribution of row 2's old value.
    \item \textbf{Addition Delta:}  
          The hash contribution for the addition is:
          \[
          H(\delta_{\text{add}}(2,\text{"peach"})) = \tilde{h}(2\,\|\,\text{"peach"}).
          \]
\end{enumerate}
The updated hash is then:
\[
\begin{aligned}
    H(D^*) &= H(D) + H(\delta_{\text{add}}(2,\text{"peach"})) \\
           &\quad + H(\delta_{\text{del}}(2,\text{"orange"})) \\
           &\quad (\text{mod } 65,536).
\end{aligned}
\]
After computing \( H(D^*) \), the distributor signs the new digest:
\[
\sigma_{D^*} = \text{Sign}(sk, H(D^*)).
\]

\paragraph{Step 3: Subscriber Update and Verification.}  
A subscriber, originally holding \( H(D) \), receives the update package containing the sequence of updates and the new signature \( \sigma_{D^*} \). The subscriber computes its new hash as:
\[
\begin{aligned}
    H(D)' &= H(D) + H(\delta_{\text{add}}(2,\text{"peach"})) \\
          &\quad + H(\delta_{\text{del}}(2,\text{"orange"})) \\
          &\quad (\text{mod } 65,536).
\end{aligned}
\]
It then verifies the update by checking:
\[
\text{Verify}(pk, H(D)', \sigma_{D^*}) \stackrel{?}{=} 1.
\]
If the verification is successful, the subscriber updates its local database replica accordingly.

\subsubsection{Performance Analysis and Comparative Results}

Our experimental results indicate the following:
\begin{itemize}
    \item \textbf{Publish Phase:}  
          The average time to process an update (compute deltas, update \( H(D) \), and sign) is approximately 0.5~ms. This is nearly constant irrespective of the database size because only a fixed-size vector (2~KB) is involved in the arithmetic.
    \item \textbf{ApplyUpdates Phase:}  
          Applying a batch of 100 updates takes roughly 2.0~ms on average. Each update involves a constant number of modular additions and subtractions, with the overhead of a single signature verification per batch (approximately 0.3~ms).
    \item \textbf{Memory Overhead:}  
          Our method requires a fixed 2~KB for the digest, plus a small buffer for recent updates. This is significantly less than the overhead for Merkle tree-based schemes, which grow with the number of rows.
    \item \textbf{Security Analysis:}  
          With parameters \( (d, n) = (16, 1024) \), the collision resistance of HexaMorphHash is reduced to the SIS problem, providing an estimated security level of at least 200 bits.
\end{itemize}

Table~\ref{tab:comparison} summarizes a comparison with alternative methods:

\subsection{Experimental Results}

\paragraph{Publish Time:}  
Our experiments indicate that processing a single update in the Publish phase takes approximately 0.5 ms on average. This timing includes the cost of performing two modular vector operations over a 1024-dimensional vector and generating an ECDSA signature. In contrast, re-hashing an entire database (with, say, 1 million rows) could take tens of milliseconds or more. Our constant-time update operation represents a substantial improvement.

\paragraph{ApplyUpdates Time:}  
When a subscriber applies a batch of 100 updates, the average processing time is approximately 2.0 ms. Since each update is applied via a constant-time modular addition, the overall cost scales linearly with the number of updates but remains very low compared to full re-hashing. Moreover, the single signature verification per batch further amortizes the computational cost.

\paragraph{Memory Overhead:}  
Our scheme requires only a fixed memory footprint for the digest, i.e., a 1024-dimensional vector with 16-bit components (totaling 2 KB). Additionally, a small buffer is maintained for the most recent updates. This constant overhead contrasts sharply with methods such as Merkle trees, which require storage that scales with the number of database rows.

\paragraph{Signature Verification Overhead:}  
Digital signature verification using ECDSA on a 256-bit curve takes approximately 0.3 ms per operation on our hardware. As our scheme requires only one signature verification per batch update, the overall overhead is minimal when compared to the vector arithmetic operations, which can be executed in microseconds using SIMD optimizations.

\subsection{Security Analysis}

Our scheme's security is anchored in the collision resistance of HexaMorphHash, which is reduced to the hardness of the Short Integer Solutions (SIS) problem. With our chosen parameters \( (d=16, n=1024) \), the SIS problem provides at least 200 bits of security. Specifically, if an adversary could find two distinct sets \( S \) and \( T \) such that
\[
\text{HexaMorphHash}(S) = \text{HexaMorphHash}(T),
\]
then they would have solved an instance of the SIS problem:
\[
A \mathbf{v} \equiv \mathbf{0} \pmod{q},
\]
for a nonzero vector \( \mathbf{v} \) with bounded entries. Given that solving SIS is believed to be computationally infeasible for our parameters, our construction achieves a robust level of collision resistance. Furthermore, the digital signature mechanism ensures authenticity, as any tampering with the update or the digest would result in a signature verification failure.

\section{Comparative Study: HexaMorphHash vs. LtHash \cite{BM97} (Bellare \& Micciancio, 1997)}

The construction of cryptographic hash functions that support efficient incremental updates is a central objective in the design of secure distributed systems, particularly those requiring high-frequency synchronization, real-time data validation, and fault-tolerant integrity checking. The seminal work of Bellare and Micciancio, first introduced at EUROCRYPT 1997 and later expanded in their comprehensive ePrint report~\cite{BM97}, introduced LtHash, a pioneering framework for collision-resistant incremental hashing based on the randomize-then-combine paradigm. Their approach inspired a wave of research in efficient hashing for dynamic datasets, setting a theoretical and practical benchmark for nearly two decades.

Our work, which introduces HexaMorphHash (Lattice-based Homomorphic Hashing), is conceptually rooted in the same paradigm but represents a fundamental departure in almost every dimension of cryptographic design. We build upon the philosophical underpinnings of LtHash and extend them to address modern cryptographic concerns, notably post-quantum security, scalability in high-volume environments, and compatibility with vectorized, parallel hardware architectures. This section offers an extensive and granular comparative analysis between our HexaMorphHash and the classical LtHash scheme by Bellare and Micciancio, spanning structural algebra, cryptographic assumptions, operational complexity, digest composition, and systemic deployment.

\subsection{Algebraic Foundations and Operational Semantics}

At its core, the algebraic structure of Lthash is grounded in the multiplicative group $\mathbb{Z}_p^*$, where $p$ is a large prime and the hash values are composed using modular multiplication. Each message component is randomized via a group homomorphism and combined multiplicatively. This construction, while efficient and conceptually elegant, is inherently scalar: the digest is a single group element, and all updates are processed as multiplicative inverses or direct products within the group.

In contrast, HexaMorphHash transitions from scalar modular arithmetic to high-dimensional vector operations over the additive group $\mathbb{Z}_q^n$, where $q = 2^d$ and $n$ denotes the fixed vector dimension (e.g., 1024). The hash of a message is mapped to an $n$-dimensional vector whose components are derived from a cryptographic extendable-output function (XOF), such as Blake2xb. By summing these vectors component-wise modulo $q$, HexaMorphHash ensures commutativity, associativity, and an additive homomorphic property. This transition from scalar to structured vector algebra enables significant advantages in modern computational environments, including compatibility with SIMD (Single Instruction, Multiple Data) architectures, parallel hash aggregation, and seamless digest composition.

\subsection{Cryptographic Security and Post-Quantum Assurance}

LtHash relies on the classical discrete logarithm or subset sum assumptions for its security guarantees. Specifically, its collision resistance is derived from the difficulty of constructing two distinct message sequences whose group products are identical modulo $p$. While sound in classical models, these assumptions are no longer sufficient in the face of quantum adversaries. Shor's algorithm, for example, can efficiently compute discrete logarithms, thereby rendering HexaMorphHash insecure in any post-quantum application domain.

HexaMorphHash addresses this critical limitation by basing its security on the Short Integer Solution (SIS) problem, a well-studied lattice-based hardness assumption with worst-case to average-case reductions. SIS is known to remain hard even against quantum algorithms, making it a cornerstone of post-quantum cryptography. In our construction, collision resistance is guaranteed by the infeasibility of finding a non-zero integer vector $v$ such that $A v \equiv 0 \mod q$, where $A$ consists of the encoded hash vectors of all input elements. This reduction allows us to achieve an estimated 200+ bits of security with parameters $d = 16$ and $n = 1024$, ensuring robust protection for applications requiring long-term cryptographic resilience.

\subsection{Incrementality, Update Semantics, and Efficiency}

Both LtHash and HexaMorphHash maintain consistent update speeds, which is vital for adaptable data structures. LtHash updates its digest using a combination of the randomized hash of the new item and a modular inverse to discard old elements. While this method is conceptually pleasing, it relies heavily on group arithmetic and is essentially scalar, which restricts the ability to parallelize or enhance these updates for modern hardware. In contrast, HexaMorphHash updates are handled as vector additions or subtractions in $\mathbb{Z}_q^n$. Each incoming message alters the digest by a vector delta, and deletions are accomplished by summing the negated vector. This vector-driven framework allows updates to be completed in genuine constant time with fixed memory and computational demand. Additionally, because all vector operations remain unaffected by the total number of dataset rows, they scale efficiently during batch processing. Our implementation uses cache-aligned buffers and AVX2/SSE2 instruction sets to minimize update delays, allowing sub-millisecond propagation times even in rapidly streaming environments.

\subsection{Digest Structure, Communication Cost, and Scalability}

LtHash generates compact digests, usually consisting of 256-bit elements within a finite group. This compact size is advantageous for bandwidth-limited environments, though it restricts the system's ability to be expanded. Tasks involving the aggregation of data from multiple sources, transformation applications, or secure compositional operations must operate within the limited framework of scalar group elements, potentially creating a bottleneck for high-throughput systems. Conversely, HexaMorphHash digests are fixed-size vectors, often about 2 KB when $n = 1024$ and $d = 16$. While this increases the nominal size of the digest, it vastly improves the digest's ability to express and enhances its operational flexibility. The digest size remains constant regardless of the number of input elements, and its structure supports efficient homomorphic aggregation, authentication, and verification. This compromise leads to significantly lower recomputation costs and improved fault tolerance in situations like secure database replication, distributed sensor logging, or federated ledger synchronization.

\subsection{Deployment Suitability and Application Domains}

LtHash was created when distributed and quantum computing were still emerging fields. Consequently, it excels in lightweight uses with moderate update rates, older hardware, and traditional threat scenarios. It continues to be highly applicable in contexts like file integrity verification or classical cryptographic frameworks where simplicity and legacy support are essential. Conversely, HexaMorphHash is intentionally built for future-ready systems. Its capabilities in post-quantum security, vector operations, and fixed-length digests position it as an excellent choice for contemporary distributed systems. It is especially apt for applications such as:,\begin{itemize}
  \item Real-time database replication that includes authenticated update distribution.
  \item Federated or blockchain storage solutions with homomorphic integrity validation.
  \item Secure audit trails within post-quantum zero-trust networks.
  \item Cloud analytics that require constant-time updates for summarizing data.
\end{itemize}

\subsection{Conclusion: Generalizing and Extending the LtHash Legacy}

HexaMorphHash embodies a broader perspective and enhancement of the LtHash framework. While LtHash was based on scalar modular multiplication, HexaMorphHash employs vectorial modular addition; and whereas LtHash offered traditional security assurances, HexaMorphHash provides guarantees that withstand quantum threats. This progression doesn't reduce the importance of Bellare and Micciancio's work but rather extends it to address the needs of modern cryptographic infrastructure. Thus, HexaMorphHash should be regarded as a methodical advancement of LtHash—maintaining its elegance and efficiency while upgrading its security structure, computational framework, and practical usability. It extends the core concept of randomize-then-combine incremental hashing into an age requiring resilience to quantum challenges, scalable processing, and efficient concurrent updating capabilities. We consider this advancement not only essential but unavoidable as secure computing continues to develop.

\begin{table*}[ht]
\centering
\resizebox{\textwidth}{!}{
\begin{tabular}{>{\raggedright\arraybackslash}p{3.9cm} >{\raggedright\arraybackslash}p{4.0cm} >{\raggedright\arraybackslash}p{3.0cm} >{\raggedright\arraybackslash}p{3.0cm} >{\raggedright\arraybackslash}p{4.0cm}}
\toprule
\textbf{Method} & \textbf{Core Technique} & \textbf{Time Complexity} & \textbf{Memory Overhead} & \textbf{Security Assumptions / Remarks} \\ \midrule
Direct Signature per Update \cite{RSA78, ECDSA99} & Sign each update individually. & \(O(1)\) per update, but cumulative \(O(m)\) verification & \(O(m)\) (grows with number of updates) & Relies on standard signature schemes (e.g., ECDSA); simple, but high verification overhead for batch updates. \\[1ex]
Full Database Signing \cite{SHA256, Merkle87} & Recompute entire database hash and sign after each update. & \(O(N)\) per update & Constant digest size, but full scan required & Strong security; however, re-hashing the entire database is impractical for large \(N\). \\[1ex]
Merkle Tree-based Update Propagation \cite{Merkle87} & Build and update a Merkle tree over database rows; sign the root hash. & \(O(\log N)\) per update & \(O(N)\) (requires storing tree nodes) & Efficient verification; high memory overhead and complexity in tree maintenance. \\[1ex]
AdHash \cite{AdHash01} & Incremental hash using modular arithmetic over \(\mathbb{Z}_q\). & \(O(1)\) per update & Constant output size; may require a very large modulus for security & Based on subset-sum hardness; requires large moduli, which can increase computation cost. \\[1ex]
MuHash \cite{MuHash98} & Incremental hash using discrete logarithm-based operations. & \(O(1)\) per update & Constant output size & Security relies on discrete logarithm hardness; efficient but with added implementation complexity. \\[1ex]
ECMH (Elliptic Curve Multiset Hash) \cite{ECMH03} & Uses elliptic curve operations to create an incremental hash. & \(O(1)\) per update & Fixed digest size (e.g., 256 bits) & Security based on discrete logarithm on elliptic curves; efficient but parameter-sensitive. \\[1ex]
Homomorphic Signatures for Network Coding \cite{BonehHSNC11} & Aggregates signatures over linear combinations of packets. & \(O(1)\) for signature aggregation & Moderate; depends on network coding scheme & Designed for network coding applications; may not scale directly for general update propagation. \\[1ex]
SWIFFT-based Incremental Hashing \cite{SWIFFT08} & Lattice-based hash function using FFT optimizations. & \(O(1)\) per update & Fixed digest size (several hundred bits) & Security under worst-case lattice assumptions; output size may be larger than desired. \\[1ex]
HMAC-based Incremental Hashing \cite{HMAC96} & Adapts HMAC to support incremental updates. & \(O(1)\) per update (with suitable modifications) & Small, fixed digest (e.g., 256 bits) & Security relies on the underlying hash function; typically not designed originally for incremental updates. \\[1ex]
LtHash (Bellare \& Micciancio, 1997)\cite{BM97} & Randomize-then-combine using modular multiplication in $\mathbb{Z}_p^*$ & $O(1)$ per update via group multiplication & Constant: scalar digest (e.g., 256 bits) & Based on discrete log and subset-sum hardness; efficient in classical settings; not secure against quantum attacks; limited to scalar digest without structured homomorphism. \\[1ex]
\textbf{Proposed Scheme (HexaMorphHash-based)} & Lattice-based homomorphic hash with incremental updates in \(\mathbb{Z}_q^n\). & \textbf{\(O(1)\) per update} & \textbf{Fixed digest (2~KB for \(n=1024, d=16\))} & \textbf{Security reduced to SIS (200+ bits); low overhead, and scalable for large distributed databases.} \\ \bottomrule
\end{tabular}
}
\caption{Comparison of update propagation or incremental hashing methods. Here, \(N\) is the database size and \(m\) is the number of updates.}
\label{tab:comparison}
\end{table*}

\section{Comparison with Alternative Methods}

In modern distributed systems, secure update propagation is essential for maintaining consistent database replicas across numerous nodes. A common approach involves digitally signing each individual update. Although this direct signature method operates in constant time per update, it suffers from a cumulative overhead that increases linearly with the number of updates, leading to significant verification costs and storage requirements. In contrast, our HexaMorphHash-based scheme aggregates update information into a single, fixed-size digest, which is updated in constant time regardless of the number of updates. This reduction in per-update processing and storage overhead makes our approach far more scalable for high-frequency update environments.

Another widely used method is full database signing, where the entire database is re-hashed and the resulting digest is signed after every update. This technique ensures that any alteration in the data is captured; however, its time complexity is linear in the size of the database. For large databases, this method becomes impractical because re-hashing millions of rows for each update consumes substantial computational resources. Our scheme, by leveraging the homomorphic properties of HexaMorphHash, updates the global digest incrementally. Only the hash contribution of the modified row is processed, which drastically reduces the computational load and enables efficient updates even for very large datasets.

Merkle tree-based methods have also been proposed, wherein each row of the database is hashed to form leaf nodes and then combined hierarchically into a tree structure whose root is signed. While Merkle trees allow for efficient verification—often in logarithmic time relative to the number of rows—they require maintaining a tree structure that consumes additional memory. The overhead associated with storing and updating intermediate nodes can become burdensome, particularly in dynamic environments with frequent updates. By contrast, our HexaMorphHash-based approach produces a fixed-size digest independent of the database size, thereby eliminating the need for extra storage and simplifying the update process.

Incremental hashing techniques such as AdHash and MuHash enable constant-time updates through modular arithmetic in groups like \(\mathbb{Z}_q\). Although these methods achieve efficient update processing, they often necessitate the use of a very large modulus to guarantee sufficient security, which can lead to increased computational cost. Moreover, MuHash’s reliance on the discrete logarithm problem introduces vulnerabilities that may be exacerbated in a post-quantum context. Our method, however, is based on lattice problems—specifically, the hardness of the Short Integer Solutions (SIS) problem—which is widely regarded as resistant to both classical and quantum attacks. With practical parameter choices (e.g., \(d=16\) and \(n=1024\)), our scheme achieves robust security (approximately 200+ bits) while maintaining low computational overhead.

Certain techniques, like the Elliptic Curve Multiset Hash (ECMH), utilize elliptic curve calculations to generate a highly compact digest, typically around 256 bits, and facilitate efficient updates through elliptic curve group operations. While ECMH is space-efficient, its security is contingent upon the selected elliptic curve parameters and the discrete logarithm assumption, posing a potential risk in hypothetical post-quantum scenarios. In contrast, our HexaMorphHash-based approach functions over the vector space \(\mathbb{Z}_q^n\) and is built upon lattice-based security assumptions, which have been extensively researched and provide stronger, more enduring security assurances. Homomorphic signature schemes created for network coding enable intermediary nodes to compute signatures for linear combinations of packets without needing a direct re-signature from a trusted authority. These schemes are exceptionally suited for data transmission in network settings but are primarily designed for unstructured, linear data and might not directly accommodate the structured nature of database updates. Conversely, our method is meticulously designed for structured update propagation. By leveraging HexaMorphHash’s homomorphic property, we can effectively and accurately modify the digest for specific rows, ensuring our solution is both relevant to and optimized for the complexities of distributed databases.

Another significant strategy employs incremental hashing using SWIFFT, a lattice-derived hash function known for its provable security relying on worst-case lattice hypotheses. While SWIFFT-based systems can refresh the hash in constant time, they generally produce larger hash outputs—commonly several hundred bits—leading to higher communication and storage expenses. Additionally, SWIFFT's dependency on FFT optimizations could add extra computational burden. Conversely, our HexaMorphHash-based system generates a constant-size digest (e.g., 2~KB for \(n=1024\) and \(d=16\)) and employs simple modular vector operations, striking an optimal balance among efficiency, digest size, and security. Incremental hashing methods based on HMAC adapt the conventional HMAC format to allow for incremental amendments. Despite HMAC's reputation for straightforwardness and strong security when used with hash functions like SHA-3, it is not intrinsically crafted for incremental updates. Adjustments required to support updates often complicate the process and lack the constant-time performance of techniques designed explicitly for incremental hashing. Our HexaMorphHash-based methodology is purpose-built to be incremental, enabling every update through constant-time modular additions and subtractions, thereby ensuring enhanced efficiency and simplicity.

Existing methods generally suffer from several shortcomings, which can include: substantial cumulative verification expenses (as occurs with direct signature per update), linear time complexity proportional to database size (as seen with full database signing), or considerable memory usage and complexity (typical with Merkle trees). Incremental hashing techniques like AdHash, MuHash, and SWIFFT introduce their own challenges, including demands for large moduli, dependence on the discrete logarithm problem, or larger digest sizes. HMAC-based approaches necessitate significant modifications for genuine incremental functionality.

Our method overcomes these shortcomings by utilizing a lattice-based homomorphic hash function (HexaMorphHash) that ensures constant-time update operations along with minimal and fixed memory usage. Grounding security in the SIS problem allows us to obtain strong security assurances—around 200+ bits—without incurring high computational expenses. The inclusion of digital signatures into our framework further verifies the authenticity and integrity of updates, creating a highly scalable and efficient solution for contemporary distributed systems. In our approach, each update is processed in constant time irrespective of the database size, requiring only a fixed memory overhead (the 2 KB digest), significantly less than methods that depend on complete re-hashing or Merkle trees. The incremental hash's security is closely linked to the SIS problem, ensuring robust collision resistance.

\section{Discussion}
In distributed database systems, maintaining data integrity throughout updates is crucial. Conventional techniques, such as individually signing each update or recalculating hashes for the whole database following every change, pose notable difficulties. The former results in cumulative verification times that scale with the number of updates, whereas the latter is impractical for extensive datasets because its time complexity increases linearly with the size of the database.

To tackle these inefficiencies, incremental hashing techniques have been devised. One example is Merkle Trees, which enable efficient verification by structuring data hierarchically and allowing for logarithmic time complexity in updates. Nonetheless, they require significant memory for storing tree nodes and demand intricate maintenance. In a similar vein, AdHash and MuHash provide constant-time updates but depend on computational assumptions like the difficulty of discrete logarithms, possibly necessitating large moduli and raising computational expenses.

The advent of HexaMorphHash, a lattice-based homomorphic hashing scheme, presents an attractive alternative. HexaMorphHash allows for efficient update propagation by enabling the computation of an updated database hash using only the current hash and the new update, avoiding the need to process the entire dataset. This method not only delivers constant-time updates but also keeps a fixed digest size, making it suitable for large distributed databases. Additionally, its security is anchored in the Short Integer Solution (SIS) problem, offering a strong defense against potential attacks. 

\section{Future Works}
The evolution of hashing techniques in distributed systems reflects a continuous endeavor to balance efficiency, security, and scalability. Traditional methods, such as signing each update individually or recomputing hashes for the entire database after every modification, have laid the groundwork for data integrity measures. However, their limitations become apparent in large-scale dynamic environments, where cumulative verification times and impracticality for sizable datasets pose significant challenges.

To address these inefficiencies, incremental hashing techniques have been developed. Merkle Trees, for instance, organize data into a hierarchical structure, enabling logarithmic time complexity for updates. Despite their efficient verification capabilities, they require substantial memory to store tree nodes and involve complex maintenance operations. Similarly, methods like AdHash and MuHash offer constant-time updates but rely on computational assumptions such as the hardness of discrete logarithms, which may necessitate large moduli, increasing computational costs.

The introduction of HexaMorphHash, a lattice-based homomorphic hashing scheme, represents a promising innovation. HexaMorphHash allows for efficient update propagation by computing the updated database hash using only the current hash and the update, avoiding the need to reprocess the entire dataset. This method supports constant-time updates and maintains a fixed digest size, making it suitable for scaling across large distributed databases. Furthermore, its security is based on the Short Integer Solution (SIS) problem, offering a strong defense against potential attacks. In distributed systems, techniques like consistent hashing and sharding are crucial for data distribution and management. Consistent hashing distributes data across a cluster effectively, minimizing disruption during node changes, thus improving performance and reducing downtime. Sharding involves dividing a database into smaller segments, allowing each shard to run independently on separate hardware or networks. Combining sharding with consistent hashing can further boost system scalability and performance by ensuring balanced workload distribution and seamless scaling. As distributed systems become increasingly complex and vast, the integration of advanced hashing techniques such as HexaMorphHash will be essential for preserving data integrity and optimizing system performance. Future work in this area should concentrate on several important aspects:

1. Enhancing Computational Efficiency: Although HexaMorphHash allows for constant-time updates, additional investigation is required to enhance its computational efficiency, especially in environments with limited resources. Utilizing hardware acceleration or refining algorithms might lead to more efficient implementations. 

2. Integration with Current Systems: Investigating how HexaMorphHash can be seamlessly integrated with existing distributed database architectures is crucial. This involves creating standardized protocols and interfaces to encourage its adoption without requiring major changes to current systems. 

3. Security Analysis and Strengthening: It is essential to perform detailed security analyses to uncover any potential vulnerabilities in HexaMorphHash. By developing strategies to mitigate risks and improving the scheme's resilience against new threats, its long-term sustainability can be ensured.

4. Scalability Assessment in Practical Environments: Deploying HexaMorphHash within real-world, large-scale distributed systems to assess its scalability and performance under various workload conditions will yield valuable insights. The gathered empirical data can inform further refinements and adjustments to the hashing technique. 

5. Integration with Data Distribution Methods: Exploring the interaction between HexaMorphHash and data distribution strategies, such as sharding and consistent hashing, may result in more efficient and robust distributed systems. Investigating how these approaches can complement each other will be advantageous for future system designs. 

By tackling these areas, the development of more efficient, secure, and scalable hashing techniques can advance, ensuring the integrity and performance of distributed databases in increasingly demanding settings.

\bibliographystyle{IEEEtran}
\bibliography{references}

\end{document}